\documentclass[a4paper,aps,pre,twocolumn,groupedaddress,showkeys,showpacs,floats,floatats,floatfix]{revtex4-1}
\usepackage[utf8]{inputenc}
\usepackage[english]{babel}
\usepackage{graphicx}
\usepackage{dcolumn} 
\usepackage{bm}
\usepackage{natbib}
\usepackage{latexsym}
\usepackage{mathrsfs}
\usepackage{amssymb}
\usepackage{amsmath}
\usepackage{amscd}
\usepackage{color}
\usepackage{pifont}
\usepackage{ulem}
\usepackage{verbatim}
\usepackage[T1]{fontenc}
\usepackage{hyperref}
\hypersetup{
  colorlinks   = true, 
  urlcolor     = black, 
  linkcolor    = red, 
  citecolor    = blue 
}
\begin{document}
\title{Exploring conservative islands using correlated and uncorrelated noise}
\author{Rafael M.~da Silva$^{1}$, Cesar Manchein$^{2}$, and Marcus W.~Beims$^{1}$}
\affiliation{$^1$Departamento de F\'\i sica, Universidade Federal do Paran\'a,
         81531-980 Curitiba, PR, Brazil}
\affiliation{$^2$Departamento de F\'\i sica, Universidade do Estado de
  Santa Catarina, 89219-710 Joinville, SC, Brazil} 
\date{\today}
%
\begin{abstract}
In this work, noise is used to analyze the penetration of regular islands in 
conservative dynamical systems. For this purpose we use the standard map 
choosing nonlinearity parameters for which a mixed phase space is present. 
The random variable which simulates noise assumes three distributions, 
namely equally distributed, normal or Gaussian, and power-law (obtained 
from the same standard map but for other parameters). To investigate the 
penetration process and explore distinct dynamical behaviors which may 
occur, we use recurrence time statistics (RTS), Lyapunov exponents (LEs) 
and the occupation rate of the phase space. Our main findings are as follows: (i) 
the standard deviations of the distributions are the most relevant quantity 
to induce the penetration; (ii) the penetration of islands induce power-law 
decays in the RTS as a consequence of enhanced trapping; (iii) for the power-law 
correlated noise an algebraic decay of the RTS is observed, even though sticky 
motion is absent; and (iv) although strong noise intensities induce an 
ergodic-like behavior with exponential decays of RTS, the largest Lyapunov 
exponent is reminiscent of the regular islands.
\end{abstract}
%
\pacs{05.45.Ac,05.45.Pq}
\keywords{Standard Map, stickiness effect, noise.} 
\maketitle

\section{Introduction}
\label{intro}

Realistic systems are always coupled to environments. Small effects 
of the environment on the system can nicely be described using random 
perturbations (noise). In Hamiltonian systems, noise induces dissipation, 
can destroy the regular dynamics, and affects transport, to mention 
few examples. The presence of noise can drastically change the dynamics and 
some regions of the phase space, unaccessible for the conservative case, 
can be reached when noise is considered. This occurs in typical mixed 
phase spaces of two-dimensional (2D) Hamiltonian systems, where the KAM tori can 
be treated as barriers in the phase space that cannot be transposed 
\cite{Ott02}. In such cases the presence of noise allows chaotic trajectories
to penetrate tori leading to new features and behaviors. In general, the 
presence of noise can modify the volume of invariant set that are scaled with 
the magnitude of the noise \cite{Mills-06}, can enhance trapping effects in 
chaotic scattering \cite{AltmannPRL}, and can change the escape rate from algebraic 
to exponential decay in scattering regions \cite{Sanjuan-08, Sanjuan-09} and from 
trajectories leaving from inside KAM curves \cite{Rodrigues-10}. In addition, 
noise affects anomalous transport phenomena such as negative mobility and 
multiple current reversals \cite{YangLong-2012}, enhances the creation and 
annihilation rates of topological defects \cite{chaos15} and postpones the onset 
of turbulence and stabilizes the three-dimensional waves which would otherwise 
undergo gradient-induced collapse \cite{chaos11}. In systems with spatiotemporal 
chaos, noise delays and advances the collapse of chaos \cite{PhysRevE.75.066209}. 
The effect of noise in one-dimensional systems has already been studied 
\cite{JPSJ-1998} as well as its influence on the transition to chaos in systems 
which undergo period-doubling cascades \cite{crutchPRL,shraimanPRL}. For this 
class of systems was demonstrated that noise can induces the escape from 
bifurcating attractors \cite{PhysRevE.80.031147}.

In this contribution we study the effects of noise on the dynamics of the 
standard map with mixed phase space adding a sequence of independent 
random variables that follows three different distributions: Gaussian, 
uniform and a power-law correlated (PLC) distribution. The motivation to 
chose such distributions is related to the context of open systems. The 
Gaussian distribution is connected to thermal baths, the uniform distribution 
due to its simplicity and the PLC distribution related to a very actual 
research area of finite and non-Markovian environments \cite{bao17,samyr14,
rosa08}, to mention a few. Our results show that, using uncorrelated noise, the 
resulting dynamics does not depend significantly on the choice of the 
distribution (Gaussian or uniform), as expected \cite{CRUTCHFIELD-82}. For a 
PLC noise, algebraic decays for the recurrence time statistics (RTS) curves were 
found, even for larger intensities of noise. The standard deviations of the 
distributions are the relevant quantity to change the dynamics. We also show 
that strong noise intensities induce an ergodic-like behaviour with exponential 
decays of RTS; however, reminiscent of the regular islands is still visible in 
the value of the Lyapunov exponent when compared to the noiseless case.

This work is organized as follows: In Sec. \ref{model} we present the 
model as well the distributions used for generate the noise. Analytical
results for the stability of central points are also presented. In Sec. 
\ref{psd} the changes in the phase space will be investigate. In Secs. 
\ref{recur} and \ref{les} the dynamics of the standard map with noise will 
be treated using RTS and Lyapunov exponent, respectively. The occupation of 
the phase space as a function of time for each case is presented in Sec. 
\ref{rate} and in Sec. \ref{conc} we summarize our main results.

\section{Standard map with noise}
\label{model}

\begin{figure}[!b]
  \centering
  \includegraphics*[width=0.97\columnwidth]{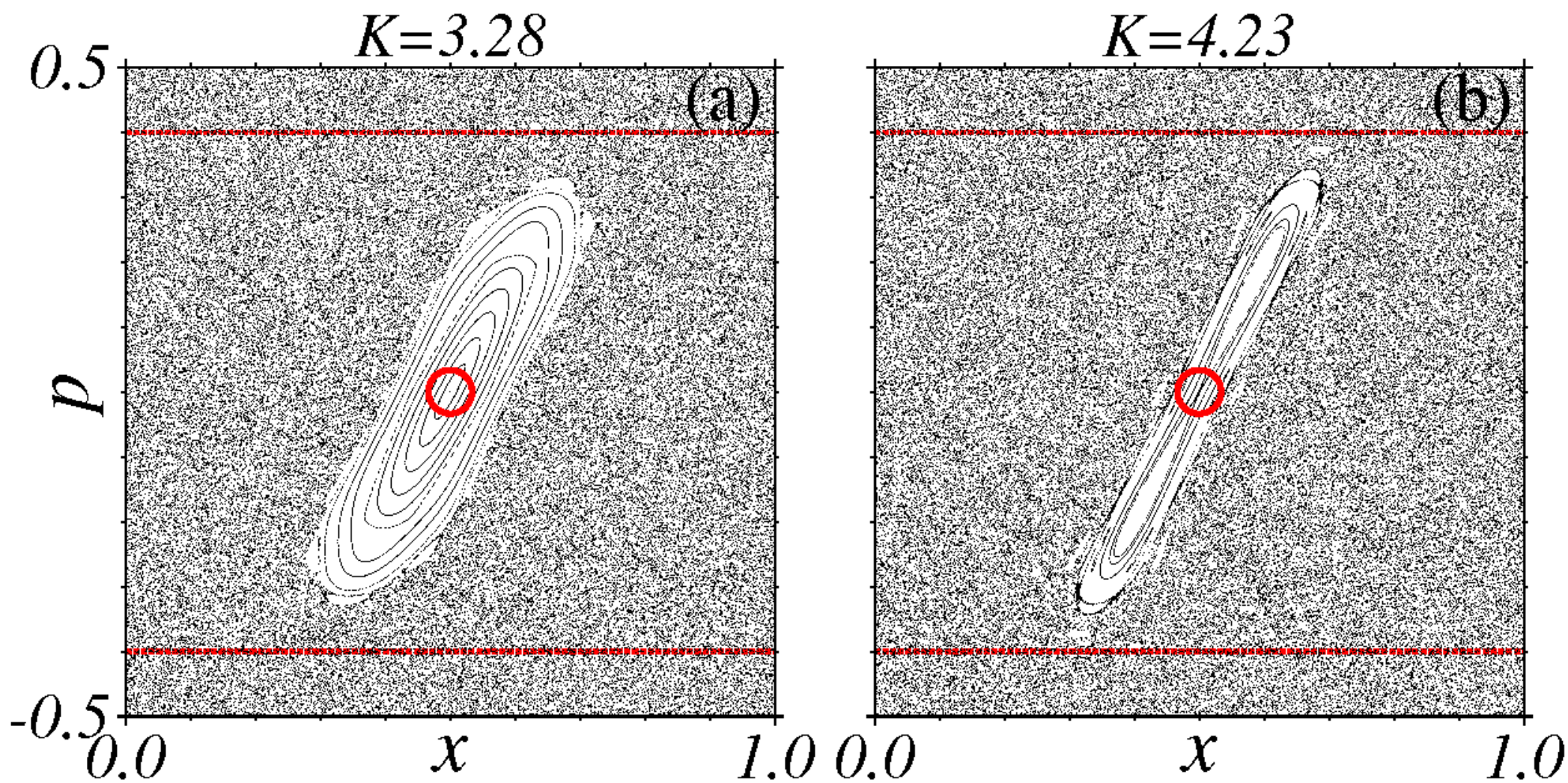}
  \caption{(Color online) Phase-space dynamics for (a) $K=3.28$, and (b) 
  $K=4.23$ using 100 initial conditions and $2\times 10^5$ iterations. 
  The fixed points (red circles) localized at $[x,p]=[1/2,0]$ in 
  the center of the figures are (a) elliptic and (b) hyperbolic.
  {Red lines define the border of the chaotic recurrence region 
  to determine the RTS, {\it i.e.} above the upper line and below the lower 
  line.}}
  \label{ps-sm}
\end{figure}

The model used in this work is the paradigmatic Chirikov-Taylor standard 
map with additive independent random variable at each time step described 
by \cite{Karney-82}
\renewcommand{\arraystretch}{1.4}
\begin{equation}
\label{stand-map}
\begin{array}{rcll}
p_{n+1}&=& p_{n} + \dfrac{K}{2\pi}\sin(2\pi \hspace{0.02cm} x_{n}) + 
\dfrac{D\xi_n}{2\pi} \hspace{0.04cm} & \hspace{0.4cm} [\mathrm{mod} \hspace{0.3cm} 1], \\
x_{n+1}&=& x_{n} + p_{n+1} & \hspace{0.4cm} [\mathrm{mod} 
\hspace{0.3cm} 1], \\
\end{array}
\end{equation}
where $x_n$ is the position at the iteration $n=0,1,2,\ldots$, and $p_n$ its 
conjugated momentum. $K$ is the nonlinear positive parameter, $\xi_n$ is the 
random variable and $D$, also a positive parameter, controls the intensity of 
$\xi_n$. The random variable was included in the above map in a distinct 
way from that proposed in \cite{Karney-82}. The parameter $K$ 
is responsible for the changes in the nonlinear dynamics, so that for larger 
values of $K$ stochasticity is obtained. The map (\ref{stand-map}) has fixed 
points at $x_1=0$, $p_1=0$ and at $x_1=1/2$, $p_1=0$. Applying the stability 
condition $|\mathrm{Tr}(\mathbf{J})| < 2$ for the trace of the Jacobian matrix 
\cite{Lichtenberg}, we find $|2\pm K| < 2$, where the upper sign corresponds 
to $x_1=0$ and the lower one to $x_1=1/2$. Solving the inequality, the point 
at $x_1=0$ is always unstable since $K$ is positive. Considering $x_1=1/2$, 
we see that for $K<4$ the fixed point is elliptic and for $K>4$ it is hyperbolic. 
These two cases are shown in Fig.~\ref{ps-sm}, using $K=3.28$ in (a) and 
$K=4.23$ in (b). For $K=3.28$ the fixed point is stable, while for $K=4.23$ 
trajectories trace a two hyperbolic branch inside the main KAM torus. For the 
values of $K$ used in this work the destroyed KAM curves form Cantor sets that 
eventually trap trajectories for a long time. This is called the sticky effect 
and is characterized by a power-law decay for the RTS curves 
\cite{Chir-Shep, Artuso, Zaslavsky2002, PhysRevLett.100.184101}.
%

\begin{figure}[!t]
  \centering
  \includegraphics*[width=0.97\columnwidth]{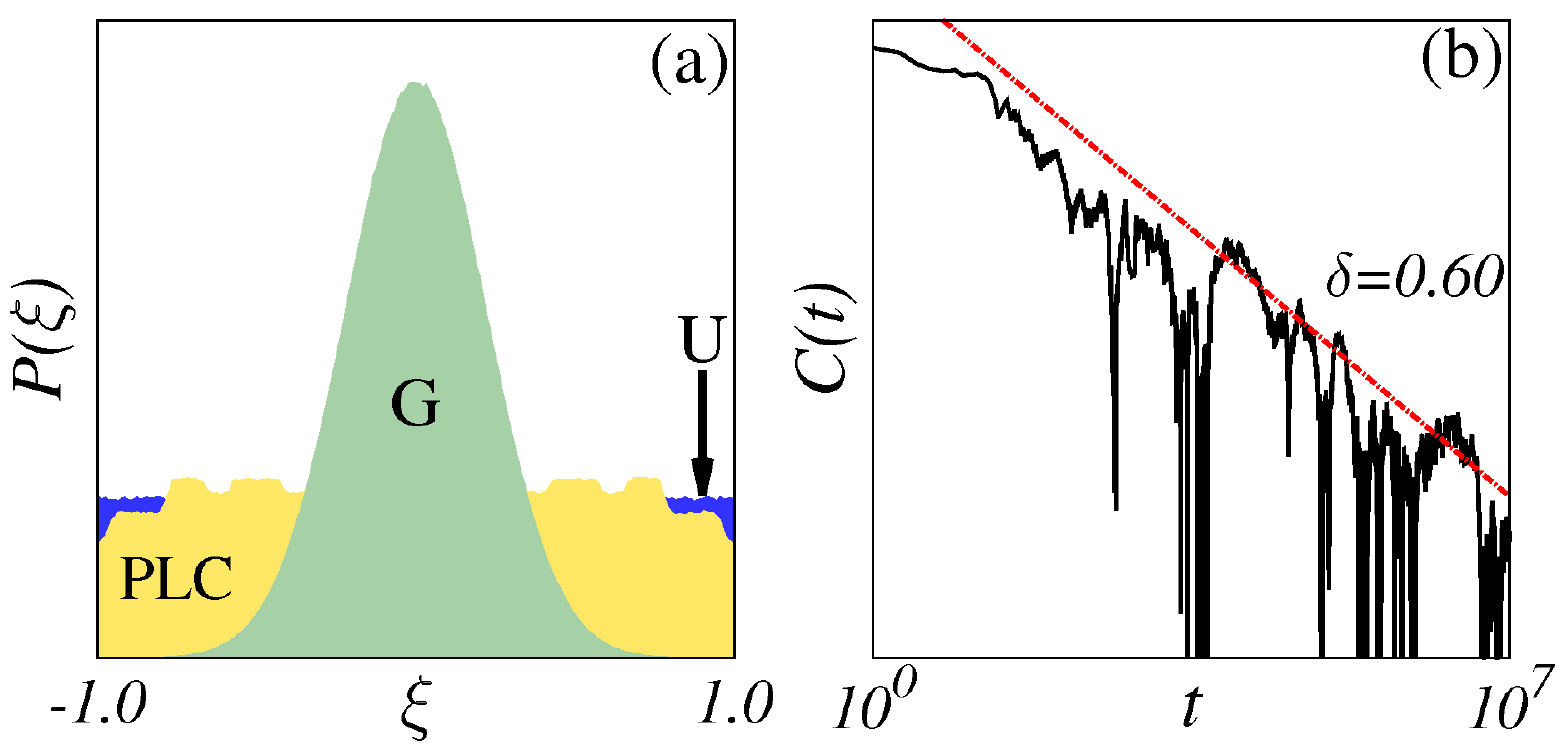}
  \caption{(Color online) (a) Probability Distributions used to generate the 
  random values $\xi$. Gaussian (green), uniform (blue) and PLC noise 
  (yellow), all cases using values inside the interval $[-1:1]$. In (b) we 
  show in logarithm scale the correlation for the variable $I_n$ of a 
  standard map for $10^7$ time iterations using $K=2.6$ that follows a 
  power-law relation $C(t)\propto t^{-\delta}$.}
  \label{dist}
\end{figure}

To generate an ensemble of uncorrelated random variables $\xi_n$ we choose: 
(i) the Gaussian (G) distribution [see green plot in Fig.~\ref{dist}(a)] with 
$\langle \xi_n\rangle=0$ and variance $0.22$, to guarantee that 
$-1\le\xi_n \le1$, and (ii) the uniform (U) distribution (see blue plot) for 
the same range, for which all values of $\xi_n$ have the same probability of 
being picked up. To obtain a correlated noise for $\xi_n$ we consider the 
standard map defined by
\renewcommand{\arraystretch}{1.4}
\begin{equation}
\label{sm-aux}
\begin{array}{rcll}
I_{n+1}&=& I_{n} + 
  \dfrac{K}{2\pi}\sin(2\pi \hspace{0.02cm} \theta_{n}), \\
\theta_{n+1}&=& \theta_{n} + I_{n+1}, \\
\end{array}
\end{equation}
\noindent where we can define the moment $I_n$ between the interval $[-1:1]$ 
and $\theta_n$ in $[0:1]$. Using $K=2.6$, case already studied before
\cite{MANCHEIN2013, RMS92}, we obtain a mixed phase space with KAM curves 
and a huge stochastic region coexisting. For a given initial condition, 
the sequence of $I_n$ and $\theta_n$ near homoclinic points generates a 
sample of values that obey a time correlated random variable 
\cite{Lichtenberg}. Doing $\xi_n = I_n$, this correlated variable is used 
here to perturb the map (\ref{stand-map}). In such cases, the time 
correlation can be determined from $C(t)=\langle \xi(t) \xi(t') \rangle$. 
For a fully chaotic phase space is expected an exponential decay: 
$C(t)\propto e^{-b\,t}$ with $t$. However, for a mixed 
phase space the correlation $C(t)$ presents a power-law tail \cite{MEISS83, 
KARNEY83, Chir-Shep, Artuso, Manchein2009}, as showed in Fig.~\ref{dist}(b) 
for $10^7$ iterations of map (\ref{sm-aux}). While 
$C(t) \propto t^{-\delta}$, the RTS curve for this mixed phase space follows 
$P(\tau) \propto \tau^{-\gamma}$, with $\gamma \sim 1.60$ \cite{RMS92}, 
while $\delta$ and $\gamma$ are related by $\delta = \gamma - 1$ 
\cite{KARNEY83,Chir-Shep,Manchein2009}. The distribution of $\xi_n = I_n$, 
which is PLC, is displayed in Fig \ref{dist}(a) by the yellow plot. 
%

\subsection{Stability condition for the central point}
\label{analitic}

Using $D\neq 0$ in map (\ref{stand-map}), no periodic orbits exist anymore.
However, for {\it one} iteration it is possible to analyze the stability of 
the ``fixed point'' under the influence of $D$. The ``fixed point'' at 
$[x_1,p_1]=[1/2,0]$ is called the {\it central point}, and note that it is not a
fixed point anymore since the noise changes its location at each iteration. Even 
though, a one-step stability analysis allows us to demonstrate that the presence of 
small noise $D\xi_n/(2\pi)$ does not change the stability condition of the 
central point. {In fact, for each iteration we are analyzing the one-step
stability of a ``fixed point'', the central point, whose location changes
anytime.} The {one-step} Jacobian matrix $\mathbf{J}_p$ of the 
standard map (\ref{stand-map}) is given by:
\renewcommand{\arraystretch}{1.4}
\begin{equation}
\mathbf{J}_p = 
\begin{bmatrix}
1 & K\cos(2\pi x_n) \\
1 & 1+K\cos(2\pi x_n)\\
\end{bmatrix}.
\label{jac1}
\end{equation}
The position of the central point is now
$p_1=0$, $x_1=-\frac{1}{(2\pi)}\arcsin[(D\xi_n)/K]$, and 
using $\cos[\arcsin(x)] = \sqrt{1-x^2}$, the Jacobian $\mathbf{J}_p$ 
becomes
\renewcommand{\arraystretch}{1.6}
\begin{equation}
\mathbf{J}_p = 
\begin{bmatrix}
1 & \pm K\sqrt{1-(D\xi_n)^2/K^2} \\
1 & 1\pm K\sqrt{1-(D\xi_n)^2/K^2} \\
\end{bmatrix}.
\label{jac2}
\end{equation}
with {eigenvalues
$h_{\pm}=\mathrm{Tr}(\mathbf{J}_p)/2\pm \sqrt{(\mathrm{Tr}(\mathbf{J}_p)^2-4)}/2$,
where} the trace {is given by}
\begin{equation}
\mathrm{Tr}(\mathbf{J}_p) = 2 \pm \sqrt{K^2-(D\xi_n)^2}.
\label{trace}
\end{equation}

{Forcing the eigenvalues of the Jacobian matrix to be 
$|h_{\pm}|<1$ implies the stability condition} 
$|\mathrm{Tr} (\mathbf{J}_p)|<2$. Applying this condition to the upper 
signal, again we have only unstable points for any value of $(D\xi_n)^2$. 
Considering the lower signal and $K=3.28$ and $K=4.23$, the stability 
condition for each case remains unaltered for $|D\xi_n| \le 1$, values 
that will be used in this work. {Therefore, all considered noise 
intensities are not strong enough to change the stability condition 
for the values of $K$ used here.}

\section{Phase-space dynamics}
\label{psd}

Plotting trajectories in phase space allows us to identify regions of chaotic
and regular motion for the standard map. When noise is included, initial
conditions chosen inside the stochastic sea can transpose the barrier of 
tori and penetrate them. In Fig.~\ref{ps-noise} the phase-space dynamics 
are shown for $K=3.28$ in (a)-(i) and $K=4.23$ in (j)-(r). The case with
Gaussian noise is displayed in the first line, (a)-(c) and (j)-(l), with 
uniform noise in the second line, (d)-(f) and (m)-(o), and with PLC noise 
in the third line, (g)-(i) and (p)-(r). Compared to  Fig.~\ref{ps-sm}(a),
for $D=10^{-5}$ only some regular trajectories inside the main torus are 
affected, as we can see in Fig.~\ref{ps-noise}(a), (d) and (g) for $K=3.28$. 
The most emblematic case is $D=10^{-3}$, for which we have a mixture of 
completely penetrated tori and other regions still unaccessible. The
increasing density of points inside the island from the case $D=0$ indicates 
that larger sticky motion is expected (this will be shown later). For these 
cases we can observe that the portion of phase space accessible for the 
trajectory depends on the distribution used. Using the uniform distribution 
with $D=10^{-3}$, the trajectories can access most of the phase space, while 
for the Gaussian distribution there are a lot of regions not visited for 
same noise intensity. This means that, using distributions for which extreme 
values of $|\xi_n|$ are most likely to occur, it is possible to access 
a larger portion of the phase space in the same time interval.
\begin{widetext}
$\quad$
\begin{figure}[!t]
  \centering
  \includegraphics*[width=1.0\columnwidth]{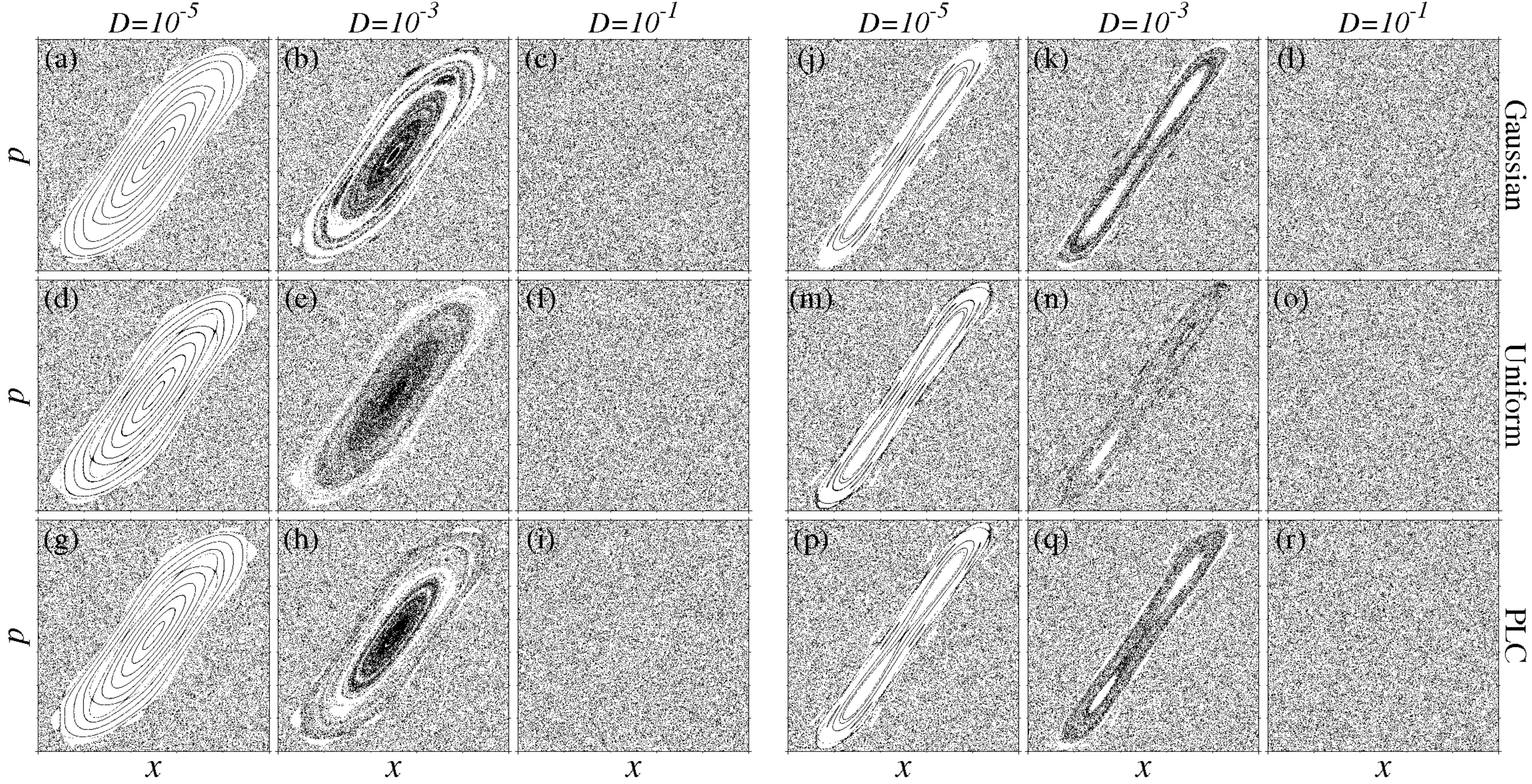}
  \caption{(Color online) Phase space of the map (\ref{stand-map}) for the 
  interval $(x_{\mbox{\tiny min}}, x_{\mbox{\tiny max}})=(0.25,0.75)$, 
  $(p_{\mbox{\tiny min}}, p_{\mbox{\tiny max}})=(-0.35,0.35)$ with $K=3.28$ 
  [(a)-(i)] and $K=4.23$ [(j)-(r)] using different intensities of noise 
  indicated above each column. The first line displays the results for 
  Gaussian noise, the second line for uniform noise and the third one for PLC 
  noise. In all simulations we used 100 initial conditions 
  and iterated the map $3\times 10^5$ times.}
  \label{ps-noise}
\end{figure}
\end{widetext}

For $D=10^{-1}$, an apparently fully chaotic motion is observed, at least 
from the phase-space dynamics analysis. However, from the analytical 
results, we know that the stable central point is still there so that 
some reminiscent of regular motion is expected. It is worthwhile mentioning 
here that for a better visualization of the phase-space dynamics we use 
shorter time iterations when compared to results from Secs.~\ref{recur},
\ref{les} and \ref{rate}. However, conclusions made above about the 
penetration of island should not be substantially changed for longer 
iterations. Besides, results from the next sections corroborate these 
findings.

\section{Recurrence Time Statistics}
\label{recur}

In this section we analyze the RTS for the system (\ref{stand-map}) using 
different intensities $D$ for each distribution. The RTS is determined 
numerically by counting the iteration times $\tau$ that the trajectory stays
outside of the recurrence region (defined inside the chaotic region). The 
existence (or not) of the sticky motion is 
recognized by the cumulative distribution $P_{\text{cum}}(\tau)$ defined by
\begin{equation}
P_{\text{cum}}(\tau) \equiv \displaystyle \sum_{\tau'=\tau}^{\infty} P(\tau'). 
\end{equation}
\noindent The quantity $P_{\text{cum}}(\tau)$ is a traditional method to 
quantify stickiness in Hamiltonian \cite{PhysRevLett.100.184101,Artuso,
AltmannKantz,Shep2010,RMS91} and conservative three-dimensional systems 
\cite{RMS92}, since events with long times $\tau$ in the RTS are associated to 
times for which the trajectory was trapped to the nonhyperbolic components of 
the phase space. {$P_{\text{cum}}(\tau)$ can be directly related with escape 
time distributions by applying the ergodic theory of transient chaos in systems 
with leaks \cite{TelPRL100, TelPRE79}}.
Although there is no general rule, algebraic decays of 
$P_{\text{cum}}(\tau)$ for at least two decades indicate the existence of sticky 
motion. When noise is added in Hamiltonian systems with mixed phase space, a 
slow additional algebraic decay of RTS curves \cite{AltmannPRL, Chir-Shep} 
and survival probability inside domains near the fixed point  
\cite{Ketzmerick2012} was observed, which means that the trapping around 
regular islands is enhanced due to trajectories that wander inside the 
islands. This result was also found in a two-dimensional conservative map coupled 
to an extra dimension without noise. In this case, trajectories remain trapped to 
the extra dimensional action variable, and for very long times no recurrence occurs, 
resulting in plateaus in the RTS curves \cite{RMS92}. In this section, our focus is 
to study the relation of enhancing trapping due to $D$ and the kind of distribution 
used, as well the influence of the stability condition of the central point of the 
standard map.

The RTS plots for $K=3.28$ are presented in Fig. \ref{rts1}(a)-(c). 
{For the recurrence box we use the chaotic region, displayed in 
Fig.~\ref{ps-sm}. It can be shown that our results are essentially 
independent on the choice of the recurrence region, as long it is located 
inside the chaotic region \cite{SalaArtusoManchein}.} 
For $D=0$ we  observe the usual algebraic decay {$P_{\text{cum}}(\tau) \propto 
\tau^{-\gamma}$} with $\gamma=1.55$, {indicating the well known sticky motion.} 
For $D \ge 10^{-2}$ no events with long recurrence times exist anymore, and the 
characteristic long tail of RTS gives place to an exponencial decay, a characteristic 
of ergodic systems. The enhanced trapping, characterized by a slower algebraic decay 
($\varepsilon=0.65$), is present for $D$ inside the interval $[10^{-5}:10^{-3}]$ 
and for all distribution. This is a consequence of the trapped motion inside de 
island from the $D=0$ case, as observed by the larger density of point in Fig. 
\ref{ps-noise}(b), (e) and (h). Since the trajectory is inside the island, there 
is a probability of occurring a sequence of $D\xi_n$ that keeps the trajectory 
trapped so that long times of recurrences are reached. The slower decay of the 
RTS curves means a decrease in the number of recurrences in this interval. 
Looking at Fig.~\ref{rts1}(c), the case for which a PLC noise was used, even 
for $D=1$ a power-law regime is obtained for the RTS curve, what does not occur 
for other distributions. The decay follows $P_{\text{cum}}(\tau) \propto 
\tau^{-\beta}$, with $\beta = 2.0$, which characterizes a superdiffusive motion 
on phase space. It is also interesting to note that for $D=10^{-3}$ and $D=10^{-2}$ 
in Fig.~\ref{rts1}(c) there is an exponential decay for long times and, increasing 
the intensity $D$, the algebraic decay is recovered. This suggests that the 
dynamics of the auxiliar map (\ref{sm-aux}) has great influence on the dynamics of 
map (\ref{stand-map}) due the relation $D\xi_n = DI_n$.

The case $K=4.23$ is displayed in Figs.~\ref{rts1}(d)-(f). By comparison with 
$K=3.28$, the enhanced trapping is not that efficient. The reason for this is 
that the region of sticky motion is smaller, as observed in  Fig. 
\ref{ps-noise}(k), (n) and (q). Besides that, there is a hyperbolic fixed point 
inside the main KAM torus forcing the trajectory to stay away from the center. 
In this case we do not find algebraic decay for RTS curves when using a PLC 
noise for larger values of $D$. Thus the sticky motion coming from the PLC is 
not able to significantly keep the sticky motion from the map (\ref{stand-map}) 
when the central point is unstable. Another important conclusion is that the 
RTS curves obtained for $K=3.28$, as well as for $K=4.23$, do not present 
relevant changes using Gaussian distribution or uniform distribution.
\begin{widetext}
$\quad$
\begin{figure}[!t]
  \centering
  \includegraphics*[width=0.99\columnwidth]{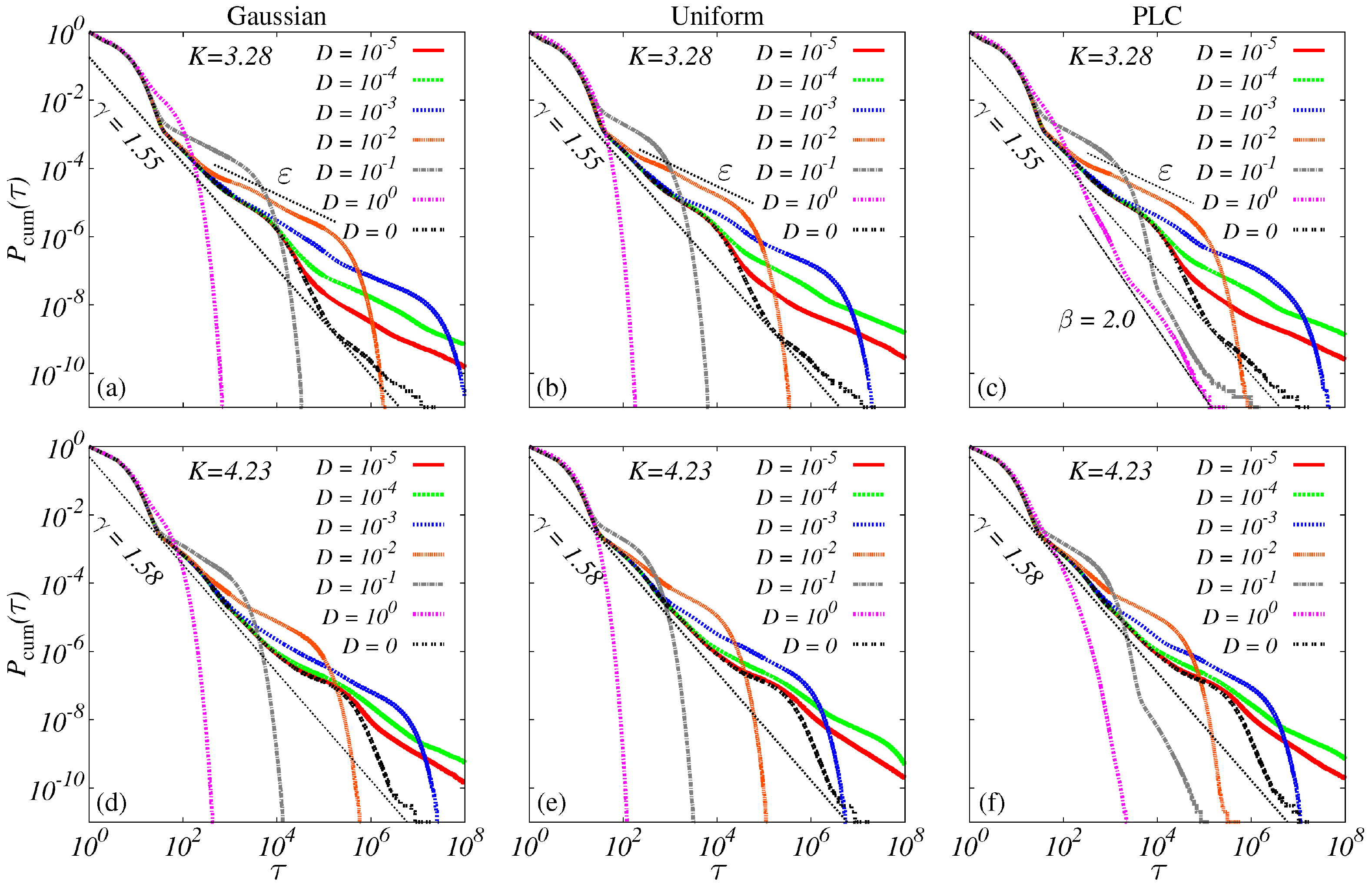}
  \caption{(Color online) Cumulative distribution $P_{\text{cum}}(\tau)$ for 
  recurrence times $\tau$ to a region in the chaotic component of the phase 
  space using Gaussian, uniform and PLC noise distributions, as indicated 
  by the title on the top of each panel with $K=3.28$ and $K=4.23$ for the 
  first and second lines of the figures, respectively. {The small algebraic 
  decay $\varepsilon=0.65$ is related to the trapped motion, or enhanced 
  trapping inside the island, and $\beta=2.0$ to the superdiffusive motion.}}
  \label{rts1}
\end{figure}
\end{widetext}

\section{Lyapunov exponent}
\label{les}

The quantity which measures the average divergence of nearby trajectories 
is the Lyapunov exponent (LE) $\lambda$, which provides a computable measure 
of the degree of stochasticity for a trajectory. A numerical method for 
computing all $2N$ LEs (namely, the Lyapunov spectrum) in a $N$ degrees of 
freedom system can be found in \cite{bggs80,wolf85}. This method includes 
the Gram-Schmidt reorthonormalization procedure. For a randomly perturbed 
system the technique to compute the Lyapunov spectrum is similar and we 
just replace the deterministic trajectory $\boldsymbol{x}$ by the perturbed 
sequence $\boldsymbol{x}^{(p)}$ \cite{CRUTCHFIELD-82,Mayer-Kress1981}. 
Considering the system studied in this work, since the noise 
$\xi_n$ is independent of $x_n$ and $p_n$, the fluctuations will not 
affect the angles between expansion and contraction directions in the 
tangent space, known as the angles between Lyapunov vectors 
\cite{beims-gallas16-1,beims-gallas16-2}, but just the probability 
distributions of variables.

\begin{figure}[!b]
  \centering
  \includegraphics*[width=0.98\columnwidth]{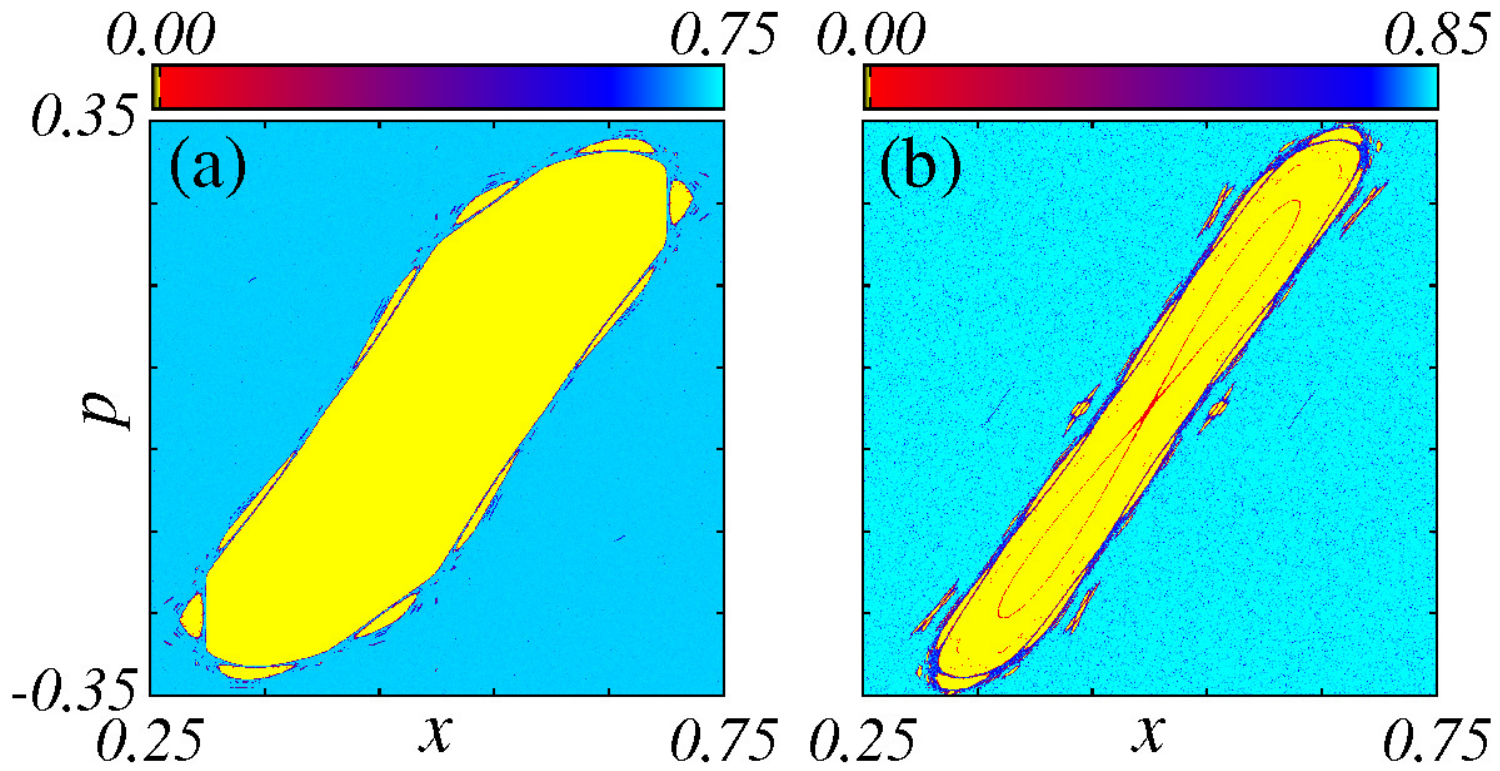}
  \caption{(Color online) The largest LE $\lambda_1$ for the map 
  (\ref{stand-map}) with $D=0$ using a grid of $10^3 \times 10^3$ initial 
  conditions for (a) $K=3.28$ and (b) $K=4.23$. For each trajectory, 
  $\lambda_1$ was calculated using $2\times 10^6$ time iterations.}
  \label{ps0-lyap}
\end{figure}

To identify changes in the dynamics of the standard map with noise, we 
divide the phase space of the map (\ref{stand-map}) for $K=3.28$ and $K=4.23$ in a 
grid with $10^3 \times 10^3$ points. Each point is an initial condition 
$[x_0,p_0]$. For trajectories starting at each combination of $[x_0,p_0]$, the 
largest LE $\lambda_1$ was determined using $2 \times 10^6$ time iterations and 
is codified by a gradient of colors in Fig. \ref{ps0-lyap} (see the color bar). 
Clearly, we observe that initial conditions inside the regular islands have 
$\lambda_1\sim 0.0$ (yellow points), with exception the unstable point in Fig. 
\ref{ps0-lyap}(b) with small positive values of $\lambda_1$ (red points). 
Initial conditions related to the chaotic trajectory have larger values of 
$\lambda_1$ (blue and cyan points).
\begin{widetext}
$\quad$
\begin{figure}[!t]
  \centering
  \includegraphics*[width=0.97\columnwidth]{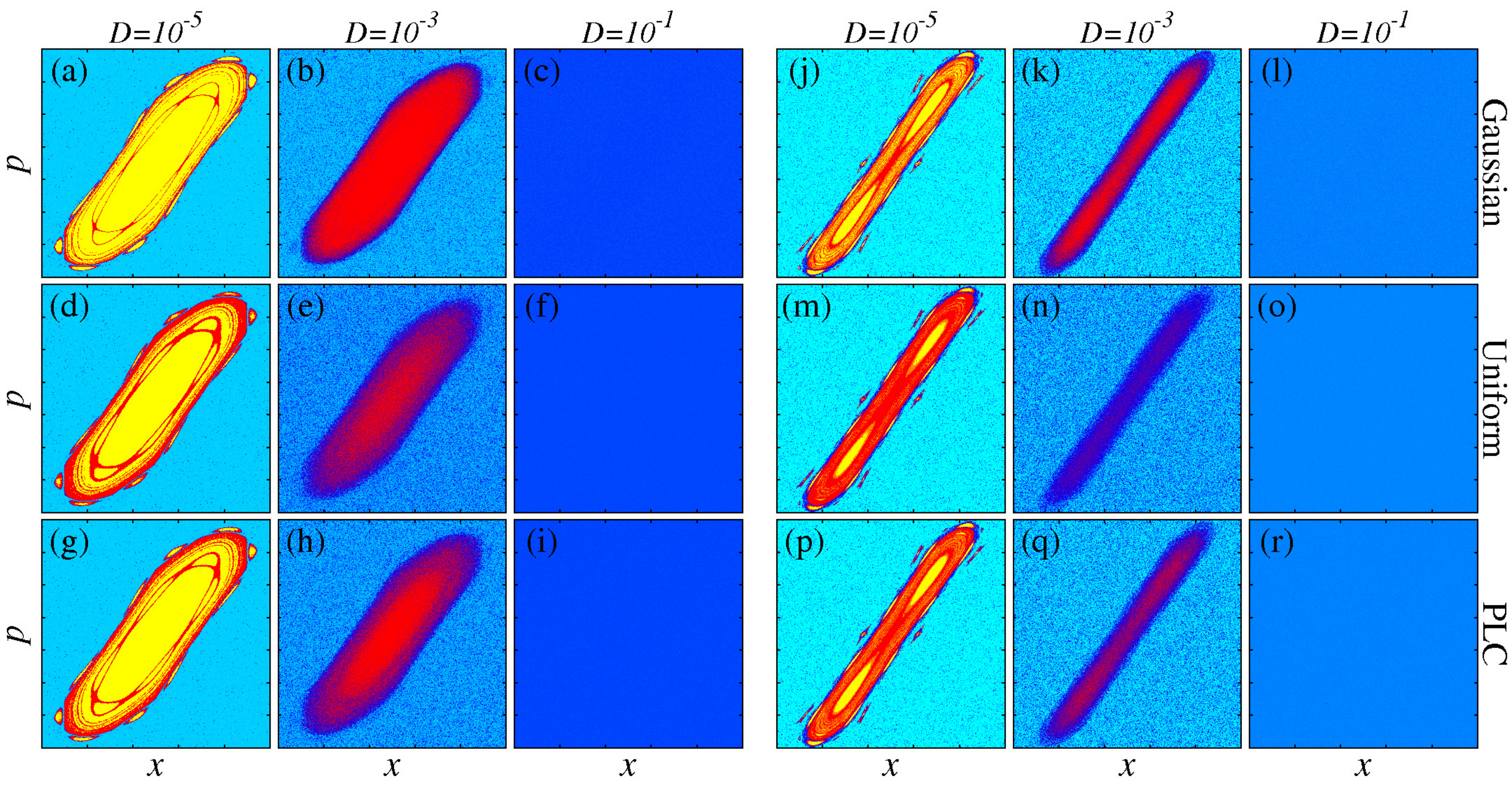}
  \caption{(Color online) The largest LE $\lambda_1$ for the map 
  (\ref{stand-map}) with $K=3.28$ [(a)-(i)], codified by the same gradient 
  color used in Fig. \ref{ps0-lyap}(a), and $K=4.23$ [(j)-(r)], codified by the 
  same gradient color used in Fig. \ref{ps0-lyap}(b). The first line was 
  obtained using the Gaussian noise, the second line using uniform noise and the 
  third line using the PLC noise. The value of $D$ for each case is indicated 
  above the correspondent column.}
  \label{ps1-lyap}
\end{figure}
\end{widetext}

If a perturbed trajectory $\boldsymbol{x}^{(p)}$ is considered, as the intensity 
$D$ of noise increases, sensitive changes can be observed in the value of 
$\lambda_1$ that are displayed in Fig. \ref{ps1-lyap}(a)-(i) for $K=3.28$ and 
(j)-(r) for $K=4.23$, using the Gaussian noise on the first line, uniform noise 
on the second line, and the PLC noise on the third. The first observation is that 
by increasing the values of $D$ the islands are penetrated and destroyed in all 
cases. For $D=10^{-1}$ (Figs. \ref{ps1-lyap}(c), (f) and (i) for $K=3.28$ and 
(l), (o) and (r) for $K=4.23$), the phase space becomes totally chaotic and the 
same value of $\lambda_1$ is obtained for all initial conditions. In other 
words, the phase space becomes ergodic-like and exponential decays are 
expected for the RTS curves, as observed in Section \ref{recur}. The relevant 
point here is to analyze how the dynamics became ergodic-like using 
distinct distributions. The Gaussian distribution does not considerably affects 
the trajectory for small values of $D$ when compared to the other distributions. 
This becomes evident when comparing the yellow region (or red) from Figs. 
\ref{ps1-lyap}(a), (d), (g), [the same for (j), (m) and (p)] that display the 
case $D=10^{-5}$ for the Gaussian, uniform, and PLC noise, respectively. The 
amount of yellow (red) points is larger (smaller) in Figs. \ref{ps1-lyap}(a) and 
(j). Besides, it is interesting to observe that when trajectories penetrate the 
islands due to noise, they tend to stay close to hyperbolic points from the tori 
transforming the dynamics more unstable (yellow $\to$ red). 

Looking at the case $D=10^{-3}$, Figs. \ref{ps1-lyap} (b), (e) and (h) for 
$K=3.28$ and (k), (n) and (q) for $K=4.23$, it is possible to note that there 
are no more initial conditions that lead to stable trajectories (yellow points). 
Using the uniform distribution [Figs. \ref{ps1-lyap}(e) and (n)] higher values 
$\lambda_1$ ($\ge 0.5$) are obtained inside the regular islands from the 
noiseless case. In addition, looking at the case $K=4.23$, an important result 
is the fast increasing of $\lambda_1$ for initial conditions around the central 
point, while for the stable case $K=3.28$ the nearby of the central point is 
kept regular for reasonable values of $D$. 

To finish this section, we would like to mention that for the ergodic-like case 
$D=10^{-1}$, already discussed above, the values of $\lambda_1$ are {\it 
smaller} than those obtained for the chaotic trajectory from $D=0$, which is 
represented as cyan. In other words, instead of increasing the values 
of $\lambda_1$ from the chaotic trajectory, the random distributions allow the 
total penetration inside the islands and traces of the regular motion are 
still visible in the asymptotic values of $\lambda_1$. Thus, the phase space of 
the ergodic-like case, which has an exponential decay of the RTS and is totally 
chaotic, still is influenced by some properties of the destroyed islands. This 
is true for correlated and uncorrelated distributions and independent of the 
presence of the stable or unstable central point.

\begin{widetext}
$\quad$
\begin{figure}[!t]
  \centering
  \includegraphics*[width=0.99\columnwidth]{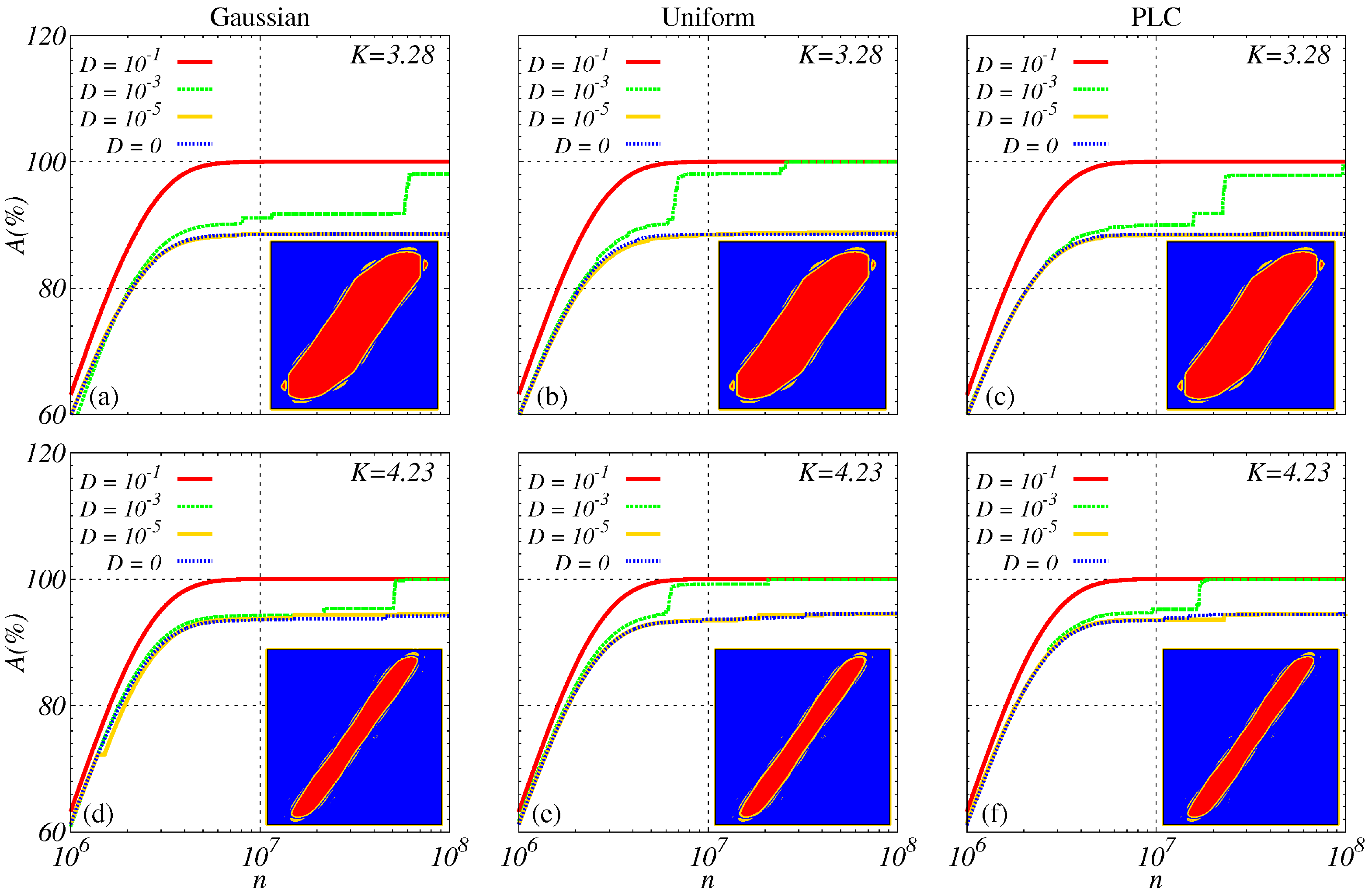}
  \caption{(Color online) {Percentage of the phase-space occupied area 
 $A(\%)$ 
  as a function of time for the three distributions and (a)-(c) $K=3.28$ and 
  (d)-(f) $K=4.23$ for some values of $D$. The inset of each case shows the 
  region visited by the trajectory. Blue points represent the visited region 
  for $D=0$, blue$+$yellow points represent the area visited for $D=10^{-5}$ 
  and blue$+$yellow$+$red is the region visited by the trajectory for 
  $D=10^{-1}$, all cases after $10^8$ time iterations.}}
  \label{ocp}
\end{figure}
\end{widetext}

\section{Phase-space occupation}
\label{rate}

The last analysis presented in this work is the occupation rate of the phase 
space as the intensity $D$ increases. In Fig. \ref{ocp}, the percentage of 
visited area $A(\%)$ of the phase space is displayed as function of the number 
of iterations $n$ for some values of $D$. Using $D=10^{-1}$ it is possible to 
access $100\%$ of the phase space after $n \approx 7 \times 10^6$ time 
iterations for any distribution (see the red curves in all panels of 
Fig.~\ref{ocp}).

For $D=10^{-3}$ the whole phase space is occupied just for $K=4.23$ [Fig. 
\ref{ocp}(d)-(f)], while for $K=3.28$ [Fig. \ref{ocp}(a)-(c)] it is possible 
only when the uniform distribution is used, and the whole phase space is visited 
after $n \approx 2.8\times 10^7$ iterations [see the green curve in Fig. 
\ref{ocp}(b)]. In this case, the abrupt increases of $A(\%)$ means that the 
penetration inside the island is also almost abrupt and not asymptotic. 

When the cases $D=0$ and $D=10^{-5}$ are compared, small differences are 
observed and we need to look at the insets that display the visited area of the 
phase space for different values of $D$. In all insets, blue points represent 
the region visited by the trajectory for $D=0$ and blue$+$yellow points 
represent the region visited for $D=10^{-5}$, both cases after $10^8$ 
iterations. Therefore the case $D=10^{-5}$ allows trajectories to access the 
high order resonances located around the main torus, what is prohibited for 
$D=0$, resulting in a small difference between the area occupied in each case. 
For some intervals of time the trajectory can be trapped in these small island 
and the visited area for $D=10^{-5}$ (yellow curves) can be smaller than the 
case $D=0$ (blue curves), as we can see in Figs. \ref{ocp}(d) for $n \approx 
1.6\times 10^{6}$ and (f) for $n \approx 1.8\times 10^7$. {It is important 
to emphasize that these results are obtained using the initial condition 
$x_0=0.159146$, $p_0=-0.470110$, localized in the chaotic sea. If other 
values are used, the curves may changed slightly but the main conclusions 
remain unaltered.}

\section{Conclusions}
\label{conc}

To summarize, we study the effects of perturbing the standard map randomly using 
an additive variable $\xi_n$ that can follow a Gaussian, a uniform and a PLC 
distribution. This last one was generated using the deterministic standard map 
with mixed dynamics. For all distributions, the RTS demonstrates that sticky 
motion is enhanced for small values of the noise intensity, namely $10^{-5}\le 
D\le 10^{-4}$. The power-law exponent characterizing this decay is $\varepsilon=
0.65$. Here the noise tends to increase hyperbolic points from the rational tori 
from the noiseless case. For intermediate values, $10^{-3}\le D\le 10^{-2}$, 
power-law decays with the same $\varepsilon$ are observed for earlier times, but 
followed asymptotically by exponential decays. This reflects the fact that larger 
noise intensities allow an earlier penetration of the island 
and the time correlation decays faster, {\it i.e}, the system 
becomes ergodic-like for earlier times when compared to smaller noise
intensities. 

For $D=10^{-1}$, all (with one exception) RTS curves decay exponentially and an 
ergodic-like motion is expected. However, the largest Lyapunov exponent is 
smaller when compared to the Lyapunov exponent from the noiseless case. This 
means that reminiscent of the sticky motion due to the destroyed islands is
still affecting the chaotic dynamics. The mentioned exception occurs when a PLC 
noise is used and $K=3.28$. In this case we found an algebraic decay with 
$\beta=2.0$, which represents a superdiffusive motion through the island. 
In fact, with these results it is pretty clear that the most relevant 
quantity to allow penetration of the island is the standard deviation of the 
distributions. 

Another issue considered in this work was the stability condition for the 
central point of the phase space. To compare the two possible conditions we 
study the influence of noise on the standard map for two different values for 
the nonlinearity parameter: $K=3.28$, for which the central point is stable, 
and $K=4.23$, for which the central point is unstable. The RTS curves from the 
Section \ref{recur} demonstrate that, in the presence of noise, the two cases 
behave similarly but the transition to stochasticity, when increasing $D$, is 
faster for the unstable case.

To finish, we would like to relate our results to higher-dimensional systems. 
Noise can be interpreted as the net effect of extra dimensions. If the dynamics 
of the extra dimensions is chaotic, the Gaussian distribution is a nice 
description of such dynamics. If the dynamics of the extra dimensions behaves 
like a conservative system with mixed phase space, then the PLC distribution 
should be adequate to describe the net effect. In this context, the global 
structure of regular tori in a generic $4$D symplectic map was analyzed 
\cite{baecker14} and, recently, the decay of RTS was studied to give a nice 
explanation about the island penetration through one extra dimension 
\cite{RMS92}.

\acknowledgments{R.M.S. thanks CAPES (Brazil) and C.M. and M.W.B. thank CNPq 
(Brazil) for financial support. The  authors  also  acknowledge computational 
support from Carlos M.~de Carvalho at LFTC-DFis-UFPR.}


\end{document}